\documentclass[12pt]{article}
\usepackage{amsmath}
\usepackage{graphicx}

\newsavebox{\SKpath}
\sbox{\SKpath}{\includegraphics[width=8.8cm]{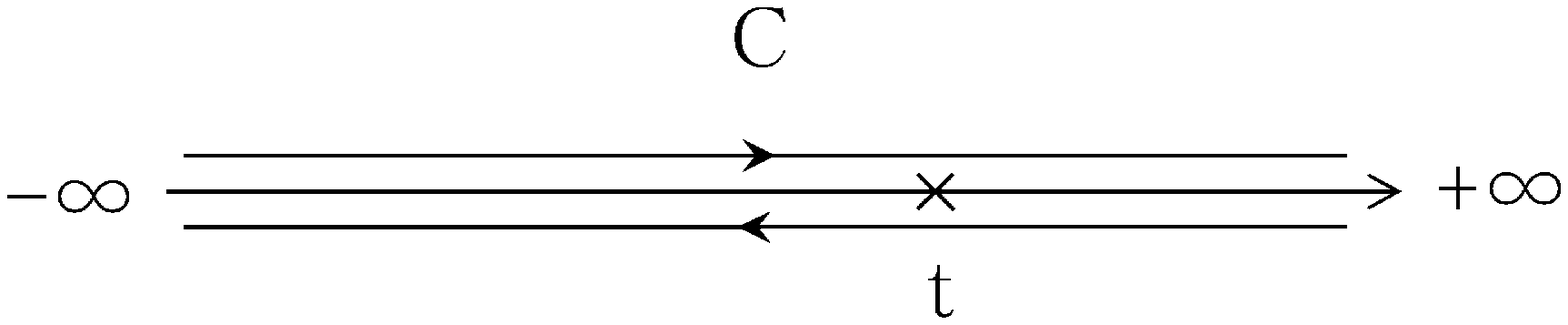}} 
\newlength{\SKpathl}
\newsavebox{\lambdaH}
\sbox{\lambdaH}{\includegraphics[width=5cm]{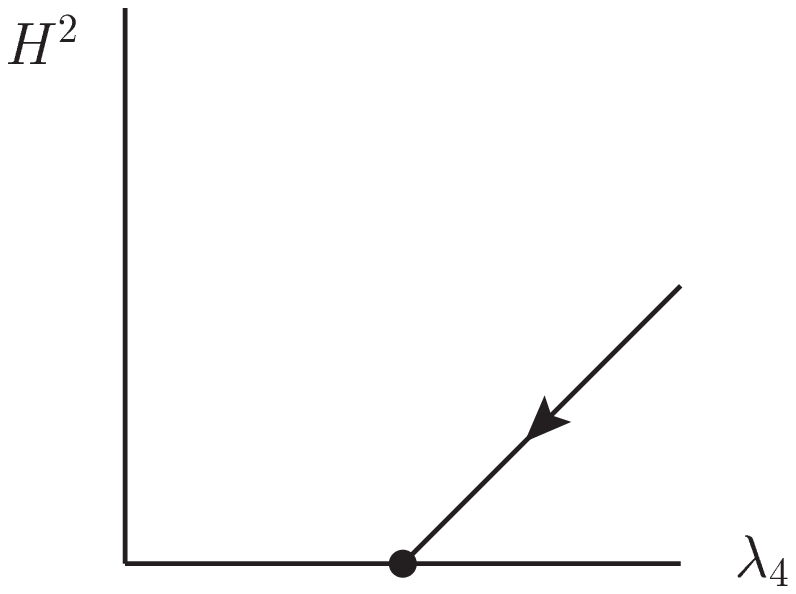}} 
\newlength{\lambdaHl}
\def\papertitlepage{\baselineskip 3.5ex\thispagestyle{empty}}
\def\preprinumber#1#2{\hfill\begin{minipage}{4.2cm} #1
        \par\noindent #2 \end{minipage}}
\allowdisplaybreaks[3]

\begin{document}

\papertitlepage
\setcounter{page}{0}
\preprinumber{KEK-TH-1704}{SNUTP14-003}
\baselineskip 0.8cm
\vspace*{6ex}

\begin{center}
{\Large\bf Time Dependent Couplings\\as Observables in de Sitter Space}
\end{center}

\begin{center}
Hiroyuki K{\sc itamoto}$^{1)}$
\footnote{E-mail address: kitamoto@snu.ac.kr}
and
Yoshihisa K{\sc itazawa}$^{2),3)}$
\footnote{E-mail address: kitazawa@post.kek.jp}\\
\vspace{5mm}
$^{1)}$
{\it Department of Physics and Astronomy}\\
{Seoul National University, Seoul 151-747, Korea}\\
$^{2)}$
{\it KEK Theory Center, Tsukuba, Ibaraki 305-0801, Japan}\\
$^{3)}$
{\it Department of Particle and Nuclear Physics}\\
{The Graduate University for Advanced Studies (Sokendai)}\\
{\it Tsukuba, Ibaraki 305-0801, Japan}\\
\end{center}

\vskip 3ex
\baselineskip = 2.5 ex

\begin{center}
{\bf Abstract}
\end{center}

We summarize and expand our investigations concerning the soft graviton effects on microscopic matter dynamics in de Sitter space. 
The physical couplings receive IR logarithmic corrections which are sensitive to the IR cut-off at the one-loop level.
The scale invariant spectrum in the gravitational propagator at the super-horizon scale is the source of the de Sitter symmetry breaking. 
The quartic scalar, Yukawa and gauge couplings become time dependent and diminish with time. 
In contrast, the Newton's constant increases with time. 
We clarify the physical mechanism behind these effects in terms of the conformal mode dynamics in analogy with 2d quantum gravity. 
We show that they are the inevitable consequence of the general covariance and lead to gauge invariant predictions.
We construct a simple model in which the cosmological constant is self-tuned to vanish due to UV-IR mixing effect. 
We also discuss phenomenological implications such as decaying Dark Energy and SUSY breaking at the Inflation era. 
The quantum effect alters the classical slow roll picture in general if the tensor-to-scalar ratio $r$ is as small as $0.01$.

{\it Keywords}: de Sitter space, time dependent couplings, Dark Energy

PACS numbers: 04.60.-m, 04.62.+v, 95.36.+x

\vspace*{\fill}
\noindent
March 2014

\newpage
\section{Introduction}\label{Introduction}
\setcounter{equation}{0}

In the present Universe, the Hubble parameter $H^2$ is so small in the unit of the Planck mass $M_P^2 = 1/ G$: 
\begin{equation}
H^2/M_P^2 = 10^{-120}. 
\label{problem}\end{equation}
Its energy scale may be related to that of neutrino mass as 
\begin{equation}
H^2M_P^2\sim m_\nu^4.
\end{equation}
It has been long thought to be exactly zero before its recent observations. 
The present Universe is well described as the space-time with a positive cosmological constant, namely de Sitter (dS) space. 
However we have not understood why the Hubble parameter (Dark Energy) is so small. 

It is likely that we cannot explain its magnitude in the present theoretical framework: relativistic quantum field theory. 
As is well known, the quantum field theory in the flat space-time is intimately connected to critical phenomena in equilibrium physics. 
In fact the space-time with a positive cosmological constant is not stationary. 
The quantum field theory in dS space may belong to nonequilibrium physics. 
In fact we need to employ the Schwinger-Keldysh formalism instead of the familiar Feynman-Dyson formalism.  

In particular, the existence of a cosmological event horizon in dS space may give rise to interesting quantum effects. 
It generates the fluctuations of the space-time metric with a scale invariant spectrum. 
Inflation theory explains the origin of large scale structures of space-time by such a mechanism.
We may need to take the fluctuations at the super-horizon scale seriously as they may eventually come back into the sub-horizon scale. 

In the Poincar\'{e} coordinate, the metric in dS space is 
\begin{align}
ds^2=-dt^2+a^2(t)d{\bf x}^2,\hspace{1em}a(t)=e^{Ht}, 
\end{align}
where the dimension of dS space is taken as $D=4$ and $H$ is the Hubble constant. 
In the conformally flat coordinate,
\begin{align}
(g_{\mu\nu})_\text{dS}=a^2(\tau)\eta_{\mu\nu},\hspace{1em}a(\tau)=-\frac{1}{H\tau}. 
\end{align}
Here the conformal time $\tau\ (-\infty <\tau < 0)$ is related to the cosmic time $t$ as $\tau=-\frac{1}{H}e^{-Ht}$. 
We assume that dS space begins at an initial time $t_i$ with a finite spatial extension. 
After a sufficient exponential expansion, dS space is well described locally by the above metric irrespective of the spatial topology. 

The metric is invariant under the scaling transformation 
\begin{align}
\tau\to C\tau,\hspace{1em}x^i\to Cx^i. 
\label{scaling}\end{align}
It is a part of the $SO(1,4)$ dS symmetry. 
The central issue is whether the dS symmetry could be broken due to IR effects. 
Note that this is a large gauge (coordinate) transformation. 
As explained shortly, we may need to introduce an IR cut-off in order to regulate the IR divergences of the propagators of the minimally coupled modes.
The IR cut-off breaks the scale invariance explicitly.
Therefore the equivalent question is to ask whether there exist IR divergences in physical observables.

Let us consider free propagators of a massless and minimally coupled scalar field $\varphi$ and a massless conformally coupled scalar field $\phi$ 
\begin{align}
\langle\varphi(x)\varphi(x')\rangle&=\frac{H^2}{4\pi^2}\big\{\frac{1}{y}-\frac{1}{2}\log y+\frac{1}{2}\log a(\tau)a(\tau')+1-\gamma\big\}, 
\label{minimally0}\end{align} 
\begin{align}
\langle\phi(x)\phi(x')\rangle&=\frac{H^2}{4\pi^2}\frac{1}{y}, 
\label{conformally0}\end{align}
where $\gamma$ is Euler's constant and $y$ is the dS invariant distance
\begin{align}
y=\frac{-(\tau-\tau')^2+({\bf x}-{\bf x}')^2}{\tau\tau'}. 
\end{align}
It should be noted that the propagator for a massless and minimally coupled scalar field has the dS symmetry breaking logarithmic term: $\log a(\tau)a(\tau')$. 

To explain what causes the dS symmetry breaking, we recall the wave function for a massless and minimally coupled field
\begin{align}
\varphi_{\bf p}(x)=\frac{H\tau}{\sqrt{2p}}(1-i\frac{1}{p\tau})e^{-ip\tau+i{\bf p}\cdot{\bf x}}. 
\end{align}
Well inside the cosmological horizon where the physical momentum $P\equiv p/a(\tau)\gg H \Leftrightarrow p|\tau|\gg 1$, 
this wave function approaches to that in Minkowski space up to a cosmic scale factor 
\begin{align}
\varphi_{\bf p}(x)\sim\frac{H\tau}{\sqrt{2p}}e^{-ip\tau+i{\bf p}\cdot{\bf x}}. 
\end{align}
On the other hand, the behavior outside the cosmological horizon $P\ll H$ is
\begin{align}
\varphi_{\bf p}(x)\sim\frac{H}{\sqrt{2p^3}}e^{i{\bf p}\cdot{\bf x}}. 
\end{align}
The IR behavior indicates that the corresponding propagator has a scale invariant spectrum. 
As a direct consequence of it, the propagator has a logarithmic divergence from the IR contributions in the infinite volume limit. 

To regularize the IR divergence, we introduce an IR cut-off $\varepsilon_0$ which fixes the minimum value of the comoving momentum 
\begin{align}
\int^H_{\varepsilon_0a^{-1}(\tau)} dP. 
\end{align}
With this prescription, more degrees of freedom come out of the cosmological horizon with cosmic evolution. 
Due to the increase, the propagator acquires the growing time dependence which spoils the dS symmetry \cite{Vilenkin1982,Linde1982,Starobinsky1982}. 
In tribute to its origin, we call the dS symmetry breaking term the IR logarithm. 
Physically speaking, $1/\varepsilon_0$ is recognized as an initial size of universe when the exponential expanding starts. 
For simplicity, we set $\varepsilon_0=H$ in (\ref{minimally0}). 
We recall here that the physical momentum $P=p/a(\tau)$ is invariant under the scaling transformation (\ref{scaling}). 
Our fundamental hypothesis is to adopt the dS invariant UV cut-off $P<\Lambda_\text{UV}$. 
In this prescription UV contributions from $H<P<\Lambda_\text{UV}$ is time independent.
In contrast, the IR contributions from $H/a(\tau)<P<H$ could grow with cosmic expansion.

Actually the metric fluctuations always contain a scale invariant spectrum just like the massless and minimally coupled scalar field. 
In dealing with the quantum fluctuation of the metric whose background is dS space, we adopt the following parametrization: 
\begin{align}
g_{\mu\nu}=\Omega^2(x)\tilde{g}_{\mu\nu},\ \Omega(x)=a(\tau)e^{\kappa w(x)}, 
\label{para1}\end{align}
\begin{align}
\det \tilde{g}_{\mu\nu}=-1,\ \tilde{g}_{\mu\nu}=(e^{\kappa h(x)})_{\mu\nu}, 
\label{para2}\end{align}
where $\kappa$ is defined by the Newton's constant $G$ as $\kappa^2=16\pi G$. 

The Lagrangian of Einstein gravity on the $4$-dimensional dS background is 
\begin{align}
\mathcal{L}_\text{gravity}&=\frac{1}{\kappa^2}\sqrt{-g}\big[R-6H^2\big]\label{gravity}\\
&=\frac{1}{\kappa^2}\big[\Omega^2\tilde{R} 
+6\tilde{g}^{\mu\nu}\partial_\mu\Omega\partial_\nu\Omega-6H^2\Omega^4\big], \notag
\end{align}
where $\tilde{R}$ denotes the Ricci scalar constructed from $\tilde{g}_{\mu\nu}$. 
In order to fix the gauge with respect to general coordinate invariance, we adopt the following gauge fixing term \cite{Tsamis1992}: 
\begin{align}
\mathcal{L}_\text{GF}&=-\frac{1}{2}a^{2}F_\mu F^\mu, \label{GF}\\
F_\mu&=\partial_\rho h_\mu^{\ \rho}-2\partial_\mu w+2h_\mu^{\ \rho}\partial_\rho\log a+4w\partial_\mu\log a. \notag
\end{align}
In this paper, the Lorentz indexes are raised and lowered by the flat metric $\eta^{\mu\nu}$ and $\eta_{\mu\nu}$ respectively. 

After decomposing the spatial part of the metric and diagonalizing the quadratic action: 
\begin{align}
h^{ij}=\tilde{h}^{ij}+\frac{1}{3}h^{kk}\delta^{ij}=\tilde{h}^{ij}+\frac{1}{3}h^{00}\delta^{ij}, 
\end{align} 
\begin{align}
X=2\sqrt{3}w-\frac{1}{\sqrt{3}}h^{00},\hspace{1em}Y=h^{00}-2w, 
\label{diagonalize}\end{align}
we find that some modes of gravity behave as the massless and minimally coupled scalar field and the other modes behaves as the massless and conformally coupled mode:    
\begin{align}
\langle X(x)X(x')\rangle&=-\langle\varphi(x)\varphi(x')\rangle, \label{minimally}\\
\langle\tilde{h}^i_{\ j}(x)\tilde{h}^k_{\ l}(x')\rangle&=(\delta^{ik}\delta_{jl}+\delta^i_{\ l}\delta_j^{\ k}-\frac{2}{3}\delta^i_{\ j}\delta^k_{\ l})\langle\varphi(x)\varphi(x')\rangle, \notag\\
\langle b^i(x)\bar{b}^j(x')\rangle&=\delta^{ij}\langle\varphi(x)\varphi(x')\rangle, \notag
\end{align}
\begin{align}
\langle h^{0i}(x)h^{0j}(x')\rangle&=-\delta^{ij}\langle\phi(x)\phi(x')\rangle, \label{conformally}\\
\langle Y(x)Y(x')\rangle&=\langle\phi(x)\phi(x')\rangle, \notag\\
\langle b^0(x)\bar{b}^0(x')\rangle&=-\langle\phi(x)\phi(x')\rangle, \notag
\end{align}
where $b$, $\bar{b}$ denote the ghost and anti-ghost fields. 

Since we focus on the dS symmetry breaking effects, we may introduce an approximation. 
We can neglect the conformally coupled modes of gravity (\ref{conformally}) since they do not induce the IR logarithm. 
In such an approximation, the following identity holds 
\begin{align}
h^{00}\simeq 2w\simeq\frac{\sqrt{3}}{2}X. 
\label{diagonalize1}\end{align}
Our aim is to investigate quantum IR effects from the minimally coupled modes of gravity (\ref{minimally}) on microscopic matter dynamics. 
To do so, we derive the effective equation of motion which takes account of  the quantum effects due to soft gravitons. 
Before investigating specific models, we review the general procedure to derive the effective equation of motion in the next section.  

\section{Schwinger-Keldysh formalism}\label{SK} 
\setcounter{equation}{0}

In this section, we review how to derive the effective equation of motion in a time dependent curved space-time. 
Let us represent the vacuum at $t\to-\infty$ as $|in\rangle$, and $t\to\infty$ as $|out\rangle$. 
In the Feynman-Dyson formalism on a flat background, 
it is presumed that $|out\rangle$ is equal to $|in\rangle$ up to a phase factor. 
On the other hand, we can't prefix $|out\rangle$ in dS space. 
The correct strategy is to evaluate vacuum expectation values (vev) with respect to $|in\rangle$: 
\begin{align}
\langle \mathcal{O}_H(x)\rangle
=\langle in| T_C[U(-\infty,\infty)U(\infty,-\infty)\mathcal{O}_I(x)]|in\rangle, 
\end{align}
where $\mathcal{O}_H$ and $\mathcal{O}_I$ denote the operators in the Heisenberg and the interaction pictures respectively. 
$U(t_1,t_2)$ is the time translation operator in the interaction picture 
\begin{align}
U(t_1,t_2)=\exp\big\{i\int^{t_1}_{t_2}d^4x\ \delta\mathcal{L}_I(x)\big\}. 
\end{align}
Here $\delta \mathcal{L}$ denotes the interaction term of the Lagrangian.  
It is crucial that the operator ordering $T_C$ specified by the following path is adopted here
\begin{align}
&\parbox{\SKpathl}{\usebox{\SKpath}}, \\
&\hspace{3em}\int_C dt = \int^\infty_{-\infty} dt_+ - \int^\infty_{-\infty} dt_-. \notag
\end{align}
We call it the Schwinger-Keldysh formalism \cite{Schwinger,Keldysh}. 
Since there are two time indices $+,-$ in this formalism, the propagator has four components
\begin{align}
\begin{pmatrix} \langle\varphi_+(x)\varphi_+(x')\rangle & \langle\varphi_+(x)\varphi_-(x')\rangle \\
\langle\varphi_-(x)\varphi_+(x')\rangle & \langle\varphi_-(x)\varphi_-(x')\rangle \end{pmatrix}
=\begin{pmatrix} \langle T\varphi(x)\varphi(x')\rangle & \langle\varphi(x')\varphi(x)\rangle \\
\langle\varphi(x)\varphi(x')\rangle & \langle\tilde{T}\varphi(x)\varphi(x')\rangle \end{pmatrix}, 
\label{4propagators}\end{align}
where $T$ denotes the time ordering and $\tilde{T}$ denotes the anti-time ordering. 

Let us introduce the external source $J_+,J_-$ for each path and evaluate
\begin{align}
Z[J_+,J_-]=
\langle in| T_C[U(-\infty,\infty)U(\infty,-\infty)
\exp\big\{i\int d^4x\ (J_+\varphi_+ -J_-\varphi_-)\big\}
]|in\rangle. 
\end{align}
The generating functional for the connected Green's functions is
\begin{align}
iW[J_+,J_-]=\log Z[J_+,J_-]. 
\label{connected}\end{align}
We define the classical field as
\begin{align}
\hat{\varphi}_A(x)=c_{AB}\frac{\delta W[J_+,J_-]}{\delta J_B(x)},\hspace{1em}A,B=+,-, 
\label{varphi}\end{align}
\begin{align}
c_{AB}=\begin{pmatrix} 1 & 0 \\ 0 & -1\end{pmatrix}. 
\end{align}
By taking the limit $J_+=J_-=J$ in (\ref{varphi}), 
we obtain the vev of $\varphi$ where the action contains the additional $J\varphi$ term 
\begin{align}
\langle \varphi(x)\rangle|_{J\varphi}=\hat{\varphi}_+(x)|_{J_+=J_-=J}=\hat{\varphi}_-(x)|_{J_+=J_-=J}. 
\label{coincident1}\end{align}
Finally, we turn off the source term $J=0$ 
\begin{align}
\langle\varphi(x)\rangle=\frac{\delta W[J_+,J_-]}{\delta J_+(x)}\Big|_{J_+=J_-=0}
=-\frac{\delta W[J_+,J_-]}{\delta J_-(x)}\Big|_{J_+=J_-=0}. 
\label{vev}\end{align}

The effective action is obtained after the Legendre transformation
\begin{align}
\Gamma[\hat{\varphi}_+,\hat{\varphi}_-]=W[J_+,J_-]-\int d^4x\ (J_+\hat{\varphi}_+ -J_-\hat{\varphi}_-), 
\end{align}
where $J_{+,-}$ are given by $\hat{\varphi}_{+,-}$ as follows
\begin{align}
J_A(x)=-c_{AB}\frac{\delta \Gamma[\hat{\varphi}_+,\hat{\varphi}_-]}{\delta \hat{\varphi}_B(x)}. 
\label{J}\end{align} 
From (\ref{coincident1}) and (\ref{J}), we obtain in the limit $\hat{\varphi}_+=\hat{\varphi}_-=\hat{\varphi}$
\begin{align}
J(x)=-\frac{\delta \Gamma[\hat{\varphi}_+,\hat{\varphi}_-]}{\delta \hat{\varphi}_+(x)}\Big|_{\hat{\varphi}_+=\hat{\varphi}_-=\hat{\varphi}}
=\frac{\delta \Gamma[\hat{\varphi}_+,\hat{\varphi}_-]}{\delta \hat{\varphi}_-(x)}\Big|_{\hat{\varphi}_+=\hat{\varphi}_-=\hat{\varphi}}. 
\label{coincident2}\end{align}
In the absence of the external source, the exact equation of motion is obtained including quantum effects
\begin{align}
\frac{\delta \Gamma[\hat{\varphi}_+,\hat{\varphi}_-]}{\delta \hat{\varphi}_+(x)}\Big|_{\hat{\varphi}_+=\hat{\varphi}_-=\hat{\varphi}}
=-\frac{\delta \Gamma[\hat{\varphi}_+,\hat{\varphi}_-]}{\delta \hat{\varphi}_-(x)}\Big|_{\hat{\varphi}_+=\hat{\varphi}_-=\hat{\varphi}}=0. 
\label{EEoM}\end{align}
The quantum equation of motion is an appropriate tool to investigate the non-equilibrium physics and curved space-times \cite{Chou,Hu1,Hu2} as we do not need to prefix the unknown vacuum state $|out\rangle$.  
Nevertheless we still need to specify the initial state $|in\rangle$. In this paper we take it to be the
Bunch-Davies vacuum.

\section{Free scalar and Dirac fields}\label{Free}
\setcounter{equation}{0}

Our interests are soft gravitational effects on the local dynamics of matter fields at the sub-horizon scale \cite{KitamotoSD,KitamotoG}. 
Although we cannot observe the super-horizon modes directly, it is possible that virtual gravitons of the super-horizon scale affect microscopic physics which are directly observable. 
We propose that observables in dS space are of this type.
In some specific models, we have derived the field equation of local matter dynamics corrected by soft gravitons at the one-loop level. 
As the first example, let us investigate a free massless conformally coupled scalar field: 
\begin{align}
S=\int \sqrt{-g}d^4x\big[-\frac{1}{2}g^{\mu\nu}\partial_\mu\phi\partial_\nu\phi-\frac{1}{12}R\phi^2\big]. 
\end{align}
We redefine the matter field as 
\begin{align}
\tilde{\phi}\equiv\Omega\phi, 
\label{frs}\end{align}
\begin{align}
S=\int d^4x\big[-\frac{1}{2}\tilde{g}^{\mu\nu}\partial_\mu\tilde{\phi}\partial_\nu\tilde{\phi}-\frac{1}{12}\tilde{R}\tilde{\phi}^2\big]. 
\label{scalar}\end{align}
This variable corresponds to the canonical quantization 
for the conformally flat metric $\tilde{g}^{\mu\nu}=\eta^{\mu\nu}$.
We believe the freedom of the tensor weight of the matter path integral
is uniquely fixed this way by the requirement of unitarity.

Up to the one-loop level, the quantum equation of motion is written as 
\begin{align}
&\partial_\mu\partial^\mu\hat{\tilde{\phi}}(x)-i\int d^4x'\ \Big[\Sigma_\text{4-pt}(x,x')+\big\{\Sigma_\text{3-pt}^{++}(x,x')-\Sigma_\text{3-pt}^{+-}(x,x')\big\}\Big]\hat{\tilde{\phi}}(x')=0,  
\label{sEoM1}\end{align}
where the self-energies from the four-point vertices and the three-point vertices are given by
\begin{align}
-i\Sigma_\text{4-pt}(x,x')=\Big[&\ \frac{1}{2}\kappa^2\partial_\mu\big\{\langle h^\mu_{\ \rho}(x)h^{\rho\nu}(x)\rangle\partial_\nu\big\}
+\frac{1}{12}\kappa^2\partial_\mu\partial_\nu\langle h^\mu_{\ \rho}(x)h^{\rho\nu}(x)\rangle \label{4-pt}\\
&+\frac{1}{24}\kappa^2\langle \partial_\mu h_{\rho\alpha}(x)\partial^\mu h^{\rho\alpha}(x)\rangle
-\frac{1}{12}\kappa^2\langle \partial_\mu h_{\rho\alpha}(x)\partial^\rho h^{\mu\alpha}(x)\rangle\Big]\delta^{(4)}(x-x'), \notag
\end{align}
\begin{align}
-i\Sigma^{AB}_\text{3-pt}(x,x')=
&\ i\kappa^2\partial_\mu\partial'_\sigma\big\{\langle (h^{\mu\nu})_A(x)(h^{\rho\sigma})_B(x')\rangle\langle\partial_\nu\tilde{\phi}_A(x)\partial'_\rho\tilde{\phi}_B(x')\rangle\big\} \label{3-pt}\\
&+\frac{i}{6}\kappa^2\partial_\mu\big\{\langle (h^{\mu\nu})_A(x)\partial'_\rho\partial'_\sigma (h^{\rho\sigma})_B(x')\rangle\langle\partial_\nu\tilde{\phi}_A(x)\tilde{\phi}_B(x')\rangle\big\} \notag\\
&+\frac{i}{6}\kappa^2\partial'_\sigma\big\{\langle \partial_\mu\partial_\nu (h^{\mu\nu})_A(x)(h^{\rho\sigma})_B(x')\rangle\langle\tilde{\phi}_A(x)\partial'_\rho\tilde{\phi}_B(x')\rangle\big\} \notag\\
&+\frac{i}{36}\kappa^2\langle \partial_\mu\partial_\nu (h^{\mu\nu})_A(x)\partial'_\rho\partial'_\sigma (h^{\rho\sigma})_B(x')\rangle\langle\tilde{\phi}_A(x)\tilde{\phi}_B(x')\rangle. \notag
\end{align}
Here the differential operators are applied after the step functions are assigned. 
This prescription corresponds with the $T^*$ product. 

In (\ref{sEoM1}), the Lorentz invariance is respected at the tree level. 
It is because the dS metric is conformally flat and the scalar field is conformally coupled.
Even for the minimally coupled case, the Lorentz invariance holds as the effective symmetry
in the sub-horizon scale since the mass term of the Hubble scale can be neglected.
The Lorentz invariance is a fundamental symmetry of the microscopic physics.
In investigating quantum IR effects, we should pay attention whether they respect the Lorentz invariance or not. 
Its possible violation could be tested stringently by experimental observations.

At the sub-horizon scale, we can neglect the derivative of the scale factor in comparison to the external momentum of the matter as 
\begin{align}
P\gg H\ \Rightarrow\ a\partial\hat{\tilde{\phi}}\gg (\partial a)\hat{\tilde{\phi}}, 
\end{align}
where $P$ denotes the external physical momentum scale of the matter. 
Furthermore the higher derivative terms arise due to UV contributions and as such they are not associated with the IR logarithm. 
We can focus on the twice differentiate term of the equation (\ref{sEoM1})
as the IR logarithms are concerned : 
\begin{align}
\log a(\tau) \partial \partial \hat{\tilde{\phi}}(x). 
\label{target}\end{align} 

It is base on a crucial fact that only the local terms contribute to the dS symmetry breaking. 
We give a detailed explanation on this fact in the next section. Here we first show the result of the first principle investigation. 
The quantum equation of motion of a scalar field including the one-loop correction from soft gravitons is evaluated as 
\begin{align}
\big\{1+\frac{3\kappa^2H^2}{32\pi^2}\log a(\tau)\big\}\partial_\mu\partial^\mu\hat{\tilde{\phi}}(x)\simeq 0. 
\label{soverall}\end{align}
We emphasize that the IR logarithm appears as an overall factor of the kinetic term. 
In this regard, soft gravitons do not spoil the Lorentz invariance at the sub-horizon scale. 

Since the derivative of $\log a(\tau)$ is negligible on the local dynamics at the sub-horizon scale, 
we can eliminate such an overall factor by the following time dependent renormalization of a scalar field:
\footnote{
For the minimally coupled massless scalar field, the wave function renormalization factor  changes due to the additional contribution from soft matter fluctuations
\begin{align}
Z_\varphi\simeq 1-\frac{\kappa^2H^2}{16\pi^2}\log a(\tau). 
\end{align}
}
\begin{align}
\tilde{\phi}\to Z_\phi\tilde{\phi},\hspace{1em}Z_\phi\simeq 1-\frac{3\kappa^2H^2}{64\pi^2}\log a(\tau). 
\label{sZ}\end{align}


It is important  to understand a physical mechanism behind this result.
We may regard the soft metric fluctuations as a slowly varying background a la \cite{Giddings2010}.
In such a view point the propagator for $\tilde{\phi}$ in the comoving momentum space 
behaves as
\begin{align}
G(x, p)=\frac{1}{\tilde{g}^{\mu\nu}p_\mu p_\nu}
=\frac{\Omega^{-2}(x)}{P_a P^a}, 
\label{mspr1}\end{align}
where the physical momentum $P_a=e^{\mu}_{~a}(x)p_{\mu}$ involves the vierbein.
Our idea is to identify the effective conformal mode dependence $e^{-2\kappa w(x)}$ in $\Omega^{-2}(x)$ as the sole source of the IR logarithms. 
In other words, we compare the magnitude of the amplitudes for the fixed physical momentum $P_a$. 
We argue that such a quantum relation is consistent with the effective Lorentz invariance. 
The conformal mode dependence can arise through the equivalence at the super-horizon scale $2w\sim h^{00}$ in (\ref{diagonalize1}). 
This idea certainly explains the time dependent wave function renormalization factor since $\langle e^{-2\kappa w}\rangle=Z_\phi^2$. 


(\ref{mspr1}) implies the following UV divergences at the coincident limit of $\tilde{\phi}$ propagator 
\begin{align}
\langle \tilde{\phi}^2(x) \rangle \sim
\int d^4p \frac{\Omega^{-2}(x)}{P_a P^a}
\sim \int d^4P \frac{\Omega^{2}(x)}{P_a P^a}
\sim \Omega^{2}(x) \Lambda_\text{UV}^2, 
\label{UVrs}
\end{align}
where we fix the UV cut-off of $P < \Lambda_{UV}$.
Since this behavior is consistent with the general covariance, 
we argue that the  (\ref{mspr1}) is required by it and hence gauge invariant.
In fact this point of view holds in more generic situations as we shall explain below.

We have performed the parallel investigation for a free massless Dirac field.  
The corresponding action is 
\begin{align}
S=\int \sqrt{-g}d^4x\ i\bar{\psi}e^\mu_{\ a}\gamma^a \nabla_\mu\psi, 
\end{align}
where $e^\mu_{\ a}$ is a vierbein and $\gamma^a$ is the gamma matrix: 
\begin{align}
\gamma^a\gamma^b+\gamma^b\gamma^a=-2\eta^{ab}. 
\end{align}
The vierbein can be parametrized as
\begin{align}
e^\mu_{\ a}=\Omega^{-1} \tilde{e}^\mu_{\ a},\hspace{1em}\tilde{e}^\mu_{\ a}=(e^{-\frac{\kappa}{2}h})_{\ a}^{\mu}.
\end{align}
In a similar way to (\ref{scalar}), we redefine the matter field such that it is canonically normalized: 
\begin{align}
\tilde{\psi}\equiv\Omega^\frac{3}{2}\psi, 
\label{frD}\end{align} 
\begin{align}
S=\int d^4x\ i\bar{\tilde{\psi}}\tilde{e}^\mu_{\ a}\gamma^a \tilde{\nabla}_\mu\tilde{\psi}. 
\label{Dirac}\end{align}

The quantum equation of motion of a Dirac field including the one-loop correction from soft gravitons is
\begin{align}
\big\{1+\frac{3\kappa^2H^2}{128\pi^2}\log a(\tau)\big\}i\gamma^\mu\partial_\mu\hat{\tilde{\psi}}(x)\simeq 0. 
\label{Doverall}\end{align}
Just like a scalar field, the IR effect from gravitons to a Dirac field preserves the Lorentz invariance. 
It can be eliminated by the following time dependent wave function renormalization of a Dirac field:
\begin{align}
\tilde{\psi}\to Z_\psi\tilde{\psi},\hspace{1em}Z_\psi\simeq 1-\frac{3\kappa^2H^2}{256\pi^2}\log a(\tau). 
\label{DZ}\end{align}
From (\ref{sZ}) and (\ref{DZ}), we can conclude that the IR logarithms due to soft gravitons can be eliminated in the free field theories 
after the wave function renormalization. 

In a similar way to the conformally coupled massless scalar field, we can reproduce the time dependent wave function renormalization of $\tilde{\psi}$ by identifying the conformal mode dependence in the propagator  as
\begin{align}
\frac{\Omega^{-1}(x)}{i\gamma^a P_a}. 
\label{mspr2}\end{align}
It is because $\langle e^{-\kappa w}\rangle= Z_\psi^2$ where we regard $P_a$ to be constant. 
This  conformal mode dependence of the propagator ensures that the UV divergences are consistent with the general covariance just like the scalar field case. 
Thus it also follows from the the fundamental symmetry as we fix the maximum value of the physical momentum $P < \Lambda_\text{UV}$.

We point out that the different choice of quantization schemes with respect to parametrization of the metric and normalization of the matter field leads to different results \cite{KitamotoPa}. 
We can reproduce the results in the existing literatures by Woodard et al. \cite{Kahya2007,Miao2005}; Giddings and Sloth \cite{Giddings2010} after taking account of these differences. 
We believe there is an advantage in our prescription.
It is consistent with unitarity since matter fields are canonically normalized.
We suspect it is the reason that we can show IR logarithmic effects  respect Lorentz invariance
of the microscopic physics.


\section{Cancellation of Non-local IR singularities}\label{Non-local}
\setcounter{equation}{0}

In investigating the dS symmetry breaking effects, we should distinguish the local and non-local IR singularities.  
When we integrate over the interaction vertices, the amplitudes with soft or collinear particles seem to induce IR singularities. 
However it is a well-known fact that such non-local IR singularities cancel after summing over degenerate states 
between real and virtual processes in QED and QCD \cite{Kinoshita1962,Lee1964}. 

We have found that the cancellation of the non-local IR singularities holds in quantum gravity on the dS background, 
at least at the one-loop level \cite{Kitamoto2013}. 
This fact indicates that only the local IR singularities contribute to the dS symmetry breaking.   

As seen in (\ref{4-pt}), at the one-loop level, the contribution from the four point vertices consists only of the local terms. 
Namely the corresponding self-energy is proportional to the delta function 
\begin{align}
\Sigma_\text{4-pt}(x,x')\propto\delta^{(4)}(x-x'). 
\end{align}  
In the coefficients of the delta function, the gravitational propagator at the coincident point 
which is left intact by differential operators $\langle h^\mu_{\ \rho}(x)h^{\rho\nu}(x)\rangle$ induces the IR logarithm. 

In contrast, the contribution from the three point vertices contains the local and the non-local terms. 
In order to extract the local terms, let us recall that the twice differentiated propagator induces the delta function. 
For example the four-times differentiated propagator of the matter field in $\Sigma_\text{3-pt}^{++}(x,x')$ induces the delta function with two differential operators: 
\begin{align}
&\partial_\mu\partial_\nu\partial_\rho\partial_\sigma\langle \tilde{\phi}_+(x)\tilde{\phi}_+(x') \rangle \label{local}\\
\to &-i\big\{
\delta_\mu^{\ 0}\delta_\nu^{\ 0}\partial_\rho\partial_\sigma+\delta_\mu^{\ 0}\delta_\rho^{\ 0}\partial_\nu\partial_\sigma
+\delta_\mu^{\ 0}\delta_\sigma^{\ 0}\partial_\nu\partial_\rho+\delta_\nu^{\ 0}\delta_\rho^{\ 0}\partial_\mu\partial_\sigma
+\delta_\nu^{\ 0}\delta_\sigma^{\ 0}\partial_\mu\partial_\rho+\delta_\rho^{\ 0}\delta_\sigma^{\ 0}\partial_\mu\partial_\nu \notag\\
&\hspace{2em}-2(\delta_\mu^{\ 0}\delta_\nu^{\ 0}\delta_\rho^{\ 0}\partial_\sigma+\delta_\mu^{\ 0}\delta_\nu^{\ 0}\delta_\sigma^{\ 0}\partial_\rho
+\delta_\mu^{\ 0}\delta_\rho^{\ 0}\delta_\sigma^{\ 0}\partial_\nu+\delta_\nu^{\ 0}\delta_\rho^{\ 0}\delta_\sigma^{\ 0}\partial_\mu)\partial_0 \notag\\
&\hspace{2em}+4\delta_\mu^{\ 0}\delta_\nu^{\ 0}\delta_\rho^{\ 0}\delta_\sigma^{\ 0}\partial_0^2
+\delta_\mu^{\ 0}\delta_\nu^{\ 0}\delta_\rho^{\ 0}\delta_\sigma^{\ 0}\partial_\alpha\partial^\alpha\big\}\delta^{(4)}(x-x'). \notag
\end{align}
We should emphasize that these local terms come from the derivative of the step function and so $\Sigma_\text{3-pt}^{-+}(x,x')$ does not contain them. 
In the coefficients of the local terms (\ref{local}), the gravitational propagator at the coincident point which is left intact by differential operators induces the IR logarithm. 

Next, let us investigate the non-local terms in the self-energy from the three point vertices. 
When we assign ${\bf p}$ as the external comoving momentum: $\hat{\tilde{\phi}}_{\bf p}(x')\propto e^{ip_\mu x'^\mu},\ p_\mu p^\mu=0$, the  non-local term is written as the following integral: 
\begin{align}
&\int d^4x'\ \big\{\Sigma_\text{3-pt}^{++}(x,x')-\Sigma_\text{3-pt}^{+-}(x,x')\big\}\big|_\text{non-local} \hat{\tilde{\phi}}_{\bf p}(x') \label{non-local}\\
=&\int d^4x'\ \theta(\tau-\tau')\big\{\Sigma_\text{3-pt}^{-+}(x,x')-\Sigma_\text{3-pt}^{+-}(x,x')\big\}\hat{\tilde{\phi}}_{\bf p}(x') \notag\\
=&\ e^{i{\bf p}\cdot{\bf x}}\int \frac{d^3p_1d^3p_2}{(2\pi)^6}\ e^{-i\epsilon\tau}\int^\tau_{\tau_i} d\tau'\ e^{i(\epsilon-p)\tau'}A({\bf p},{\bf p}_1,{\bf p}_2,\tau,\tau')
(2\pi)^3\delta^{(3)} ({\bf p}_1+{\bf p}_2-{\bf p}) -\text{(h.c.)}. \notag
\end{align}
Here ${\bf p}_1$ and ${\bf p}_2$ are respectively the comoving momenta of the intermediate scalar and gravitational fields. 
Furthermore we have introduced the total energy of intermediate particles as $\epsilon\equiv p_1+p_2$. 
$A({\bf p},{\bf p}_1,{\bf p}_2,\tau,\tau')$ merely denotes the coefficient of the oscillating phase
factor. 

In the process with a soft or collinear particle, the total energy is close to $p$: 
\begin{align}
p_1\sim 0\ \text{or}\ p_2\sim 0\ \text{or}\ {\bf p}_1,{\bf p}_2\parallel {\bf p}\ \Rightarrow\ \epsilon\sim p. 
\end{align}
In such a process, the contribution from the negatively large conformal time region is dominant since the frequency of the integrand vanishes. 
In other words, the integral (\ref{non-local}) is sensitive to the initial time $\tau_i$. We call it the non-local IR singularity.  
Strictly speaking, the non-local IR singularity comes from the integral of $\Sigma_\text{3-pt}^{-+}(x,x')$ but not of $\Sigma_\text{3-pt}^{+-}(x,x')$. 

By considering the off-shell effective equation of motion: $p_\mu p^\mu\not =0$, we can avoid the non-local IR singularity. 
It is because the lower bound of the total energy is given by the virtuality: 
\begin{align}
\epsilon^2-(p^0)^2>p_\mu p^\mu. 
\end{align}
In the subsequent discussion, we define a physical quantity which 
gives a proper interpretation into the on-shell limit of the off-shell effective equation of motion. 

It is natural to conjecture that  the non-local IR singularities cancel after summing over degenerate states between real and virtual processes also in dS space, like QED or QCD. 
In order to confirm the conjecture, we adopt the Kadanoff-Baym method \cite{Kitamoto2013}. 
The method is valid when the external momentum is at the sub-horizon scale as a particle description holds. 
In contrast to the investigation by the effective equation of motion, we investigate the 
Schwinger-Dyson equation of the two point function in this method.  
So we can systematically obtain the on-shell term and the off-shell term. 
The two point function depends on two time variables $\tau_1,\tau_2$.
We decompose them as $\tau_c\equiv (\tau_1+\tau_2)/2\gg \Delta\tau\equiv \tau_1-\tau_2$. 

We see the following integrals in the investigation of the non-local terms by the Kadanoff-Baym method: 
\begin{align}
&\int d^4x'\ \theta(\tau_1-\tau')\big\{\Sigma_\text{3-pt}^{-+}(x_1,x')-\Sigma_\text{3-pt}^{+-}(x_1,x')\big\}G_{\bf p}^{-+}(x',x_2) \label{non-local2}\\
&-\int d^4x'\ \theta(\tau_2-\tau')\Sigma_\text{3-pt}^{-+}(x_1,x')\big\{G_{\bf p}^{-+}(x',x_2)-G_{\bf p}^{+-}(x',x_2)\big\}. \notag
\end{align}
Here we have introduced the Fourier transformation of the matter propagator: $G_{\bf p}^{-+}(x_1,x_2)\propto e^{-ip_\mu(x_1-x_2)^\mu},\ p_\mu p^\mu=0$. 
There is a counter part to the first integral in the effective equation of motion (\ref{non-local}). 
After performing the integrations other than over the total energy, each integral is written as
\begin{align}
&\int d^4x'\ \theta(\tau_1-\tau')\big\{\Sigma_\text{3-pt}^{-+}(x_1,x')-\Sigma_\text{3-pt}^{+-}(x_1,x')\big\}\big|_\text{non-local}G_{\bf p}^{-+}(x',x_2) \label{on}\\
\sim &-iA'e^{i{\bf p}\cdot({\bf x}_1-{\bf x}_2)}e^{-ip(\tau_1-\tau_2)}\int^\infty_p d\epsilon\ \frac{1}{\epsilon-p}, \notag 
\end{align} 
\begin{align}
&-\int d^4x'\ \theta(\tau_2-\tau')\Sigma_\text{3-pt}^{-+}(x_1,x')\big\{G_{\bf p}^{-+}(x',x_2)-G_{\bf p}^{+-}(x',x_2)\big\} \label{off}\\
\sim&+i A'e^{i{\bf p}\cdot({\bf x}_1-{\bf x}_2)}\int^\infty_p d\epsilon\ e^{-i\epsilon(\tau_1-\tau_2)}\frac{1}{\epsilon-p}. \notag
\end{align}
In terms of the characteristic frequency of the oscillating phase factor, we call (\ref{on}) with $p$ the on-shell term and (\ref{off}) with $\epsilon$ the off-shell term. 
The on-shell and off-shell terms have the non-local IR singularities at $\epsilon\sim p$ whose coefficients $\mp iA'$ are identical except for opposite signs. 
If we naively define the on-shell and off-shell terms such as (\ref{on}), (\ref{off}), each term has an IR divergence and the IR cut-off seems to be given by the inverse of the initial time: $\int_{p+|1/\tau_i|}d\epsilon$.  

Physically speaking, any experiment has a finite energy resolution of observation $\Delta \epsilon$. 
We may divide the integration region of the off-shell term as 
\begin{align}
\int^\infty_p d\epsilon\ e^{-i\epsilon(\tau_1-\tau_2)}=\int^\infty_{p+\Delta \epsilon} d\epsilon\ e^{-i\epsilon(\tau_1-\tau_2)}+\int^{p+\Delta \epsilon}_p d\epsilon\ e^{-i\epsilon(\tau_1-\tau_2)}. 
\end{align}
Within the energy resolution, we cannot distinguish the off-shell term from the on-shell term 
\begin{align}
\int^{p+\Delta \epsilon}_p d\epsilon\ e^{-i\epsilon(\tau_1-\tau_2)}\sim e^{-ip(\tau_1-\tau_2)}\int^{p+\Delta \epsilon}_p d\epsilon. 
\end{align}
Thus we need to redefine the on-shell term by transferring the contribution of the off-shell term within the energy resolution $p<\epsilon<p+\Delta\epsilon$: 
\begin{align}
&-iA'e^{i{\bf p}\cdot({\bf x}_1-{\bf x}_2)}e^{-ip(\tau_1-\tau_2)}\Big\{\int^\infty_p d\epsilon-\int^{p+\Delta\epsilon}_p\Big\}\ \frac{1}{\epsilon-p} \label{on2}\\
=&-iA'e^{i{\bf p}\cdot({\bf x}_1-{\bf x}_2)}e^{-ip(\tau_1-\tau_2)}\int^\infty_{p+\Delta \epsilon} d\epsilon\ \frac{1}{\epsilon-p}. \notag
\end{align}
The remaining contribution is the well-defined off-shell term: 
\begin{align}
+iA'e^{i{\bf p}\cdot({\bf x}_1-{\bf x}_2)}\int^\infty_{p+\Delta \epsilon} d\epsilon\ e^{-i\epsilon(\tau_1-\tau_2)}\frac{1}{\epsilon-p}. 
\end{align}
We have found that there is no IR divergence after the redefinition. 
Since the energy resolution of observation is at the sub-horizon scale $\Delta\epsilon\sim 1/\Delta\tau\gg |1/\tau_c|>|1/\tau_i|$, 
the IR cut-off is given by not the inverse of the initial time $|1/\tau_i|$ but the energy resolution $\Delta\epsilon$. 
The non-local IR effect respects the dS symmetry as we fix the physical scale of the energy resolution $\Delta E\equiv \Delta\epsilon H|\tau_c|$. 

Furthermore we can identify the mechanism how the cancellation takes place. 
Since the zero frequency process is contained in the common integrand $\Sigma_\text{3-pt}^{-+}(x_1,x')G^{-+}_{\bf p}(x',x_2)$, 
the integral of the non-local terms (\ref{non-local2}) is evaluated as 
\begin{align}
\Big\{\int^{\tau_1}_{\tau_i}d\tau'-\int^{\tau_2}_{\tau_i}d\tau'\Big\}\int d^3x'\ \Sigma_\text{3-pt}^{-+}(x_1,x')G_{\bf p}^{-+}(x',x_2). 
\end{align}
Manifestly the initial time dependences (dS symmetry breaking effects) are canceled. 

We summarize this section. 
As for the non-local IR singularities which originate in the integrations over the vertices including soft or collinear particles, 
they cancel after summing over degenerate states between real and virtual processes like QED and QCD. 
Once the non-local terms are expressed by physical scales such as $P$, $\Delta E$, they do not contribute to the dS symmetry breaking. 
In the coefficients of the local terms, the gravitational propagator at the coincident point breaks the dS symmetry. 
So we can conclude that only through the local terms, soft gravitons contribute to the dS symmetry breaking. 

We have confirmed the cancellation of the non-local IR singularities in the two point function but not in the multi-point functions. 
In the next section, we investigate soft gravitational effects in interacting field theories.  We assume that the cancellation holds also in the multi-point functions. 
The assumption is reasonable since the cancellation originates from the fact that the total spectrum weight is preserved.  
In this regard, it may be a universal phenomenon as far as field theoretic models are consistent with unitarity. 

\section{Gauge theory, $\phi^4$ theory and Yukawa theory}\label{Interaction}
\setcounter{equation}{0}

In this section, we investigate soft gravitational effects in interacting field theories. 
For example, let us consider the gauge theory 
\begin{align}
S_\text{gauge}&=\int\sqrt{-g}d^4x\big[-\frac{1}{4e^2}g^{\mu\rho}g^{\nu\sigma}F_{\mu\nu}^a F_{\rho\sigma}^a
+i\bar{\psi}e^\mu_{\ \alpha}\gamma^\alpha D_\mu\psi\big], \label{gauge}\\
&=\int d^4x\big[-\frac{1}{4e^2}\tilde{g}^{\mu\rho}\tilde{g}^{\nu\sigma}F_{\mu\nu}^a F_{\rho\sigma}^a
+i\bar{\tilde{\psi}}\tilde{e}^\mu_{\ \alpha}\gamma^\alpha \tilde{D}_\mu\tilde{\psi}\big]. \notag
\end{align}
We have redefined the Dirac field as $\tilde{\psi}=\Omega^\frac{3}{2}\psi$ in the second line of (\ref{gauge}).   
Here the gauge group is generic and the Dirac field could be in any representation. 

Up to the one-loop level and $\mathcal{O}(\log a(\tau))$, the bosonic part of the quantum equation of motion is written as 
\begin{align}
&\frac{1}{e^2}(\hat{D}_\mu \hat{F}^{\mu\nu})^a(x)
+\kappa^2\big\{\langle (h^{\mu\rho})_+(x)(h^{\nu\sigma})_+(x)\rangle
+\frac{1}{2}\langle (h^{\mu\alpha})_+(x)(h_\alpha^{\ \rho})_+(x)\rangle\eta^{\nu\sigma} \label{F1}\\
&\hspace{9em}+\frac{1}{2}\eta^{\mu\rho}\langle (h^{\nu\alpha})_+(x)(h_\alpha^{\ \sigma})_+(x)\rangle\big\}\frac{1}{e^2}(\hat{D}_\mu \hat{F}_{\rho\sigma})^a(x). \notag
\end{align}
Here we have neglected the differentiated gravitational propagator which does not induce the IR logarithm. 
From (\ref{minimally}) and (\ref{diagonalize1}), the bosonic part is evaluated as
\begin{align}
\frac{1}{e^2}\big\{1+\frac{3\kappa^2H^2}{8\pi^2}\log a(\tau)\big\}(\hat{D}_\mu \hat{F}^{\mu\nu})^a(x). 
\end{align}
The result preserves the gauge symmetry manifestly 
and indicates that the coupling of the gauge interaction decreases with cosmic expansion 
\begin{align}
e_\text{eff}\simeq e\big\{1-\frac{3\kappa^2H^2}{16\pi^2}\log a(\tau)\big\}. 
\label{EGc}\end{align}
It is because there is no wave function renormalization for the classical gauge field in the background gauge.
The Lorentz invariance is also preserved and the velocity of light is not renormalized just like the massless scalar and Dirac fields. 

We further investigate the fermionic current $\hat{\bar{\tilde{\psi}}}\gamma^\nu t^a\hat{\tilde{\psi}}$ in the quantum equation of motion.
Up to the one-loop level, this term is evaluated as 
\begin{align}
\big\{1+\frac{3\kappa^2H^2}{128\pi^2}\log a(\tau)\big\}\hat{\bar{\tilde{\psi}}}(x)\gamma^\nu t^a\hat{\tilde{\psi}}(x). 
\label{nr}\end{align}
We recall here that the IR logarithm due to soft gravitons modifies the kinetic term of the Dirac field. 
We have shown that this change of the kinetic term can be absorbed by the wave function renormalization (\ref{DZ}). 
It should be noted that the quantum correction in (\ref{nr}) can also be absorbed by the identical wave function renormalization. 
We therefore conclude that the fermionic current is not renormalized by soft gravitons after the wave function renormalization in accord with the gauge invariance.

Let us move on to the investigation of soft gravitational effects  in $\phi^4$ and Yukawa theories. 
Since the coupling constants are dimensionless, 
$\sqrt{-g}$ can be absorbed by the field redefinition $\tilde{\phi}=\Omega\phi,\ \tilde{\psi}=\Omega^\frac{3}{2}\psi$
\begin{align}
\delta \mathcal{L}_4=-\frac{\lambda_4}{4!} \tilde{\phi}^4,  
\label{quartic}\end{align}
\begin{align}
\delta \mathcal{L}_Y=-\lambda_Y\tilde{\phi}\bar{\tilde{\psi}}\tilde{\psi}. 
\label{Yukawa}\end{align}
After the wave function renormalization (\ref{sZ}), (\ref{DZ}), 
the interaction terms are renormalized as 
\begin{align}
\delta \mathcal{L}_4=-\frac{\lambda_4}{4!} Z_\phi^4\tilde{\phi}^4,  
\label{quartic1}\end{align}
\begin{align}
\delta \mathcal{L}_Y=-\lambda_Y Z_\phi Z_\psi^2\tilde{\phi}\bar{\tilde{\psi}}\tilde{\psi}. 
\label{Yukawa1}\end{align} 

In addition to these wave function renormalization factors, soft gravitons dressing the interaction vertices modify the coupling constants. 
Considering the both contributions, we have found that the effective couplings of $\phi^4$ and Yukawa interactions decrease with cosmic expansion under the influence of soft gravitons 
\begin{align}
(\lambda_4)_\text{eff}\simeq\lambda_4\big\{1-\frac{21\kappa^2H^2}{16\pi^2}\log a(\tau)\big\}, 
\label{4ECC}\end{align}
\begin{align}
(\lambda_Y)_\text{eff}\simeq \lambda_Y\big\{1-\frac{39\kappa^2H^2}{128\pi^2}\log a(\tau)\big\}. 
\label{YECC}\end{align}
As mentioned in the previous section, we have assumed that only the local terms contribute to the dS symmetry breaking also in multi-point functions. 
\footnote{
If the minimally coupled massless scalar is involved in the interaction, the couplings evolve as
\begin{align}
(\lambda_4)_\text{eff}&\simeq \lambda_4\big\{1-\frac{25\kappa^2H^2}{16\pi^2}\log a(\tau)\big\}, 
\end{align}
\begin{align}
(\lambda_Y)_\text{eff}\simeq \lambda_Y\big\{1-\frac{41\kappa^2H^2}{128\pi^2}\log a(\tau)\big\}. 
\end{align}
}

Although these results are obtained through deriving the quantum equation of motion, 
we can simply reproduce them by using the effective propagators. 
Let us investigating the following products of them for the four point Green's function:
\begin{align}
\langle G(x, p_1)G(y, p_2)G(v, p_3)G(z, p_4)\rangle
\sim  \langle\frac{\Omega^{-2}(x)\Omega^{-2}(y)\Omega^{-2}(v)\Omega^{-2}(z)}
{P_1^2P_2^2P_3^2P_4^2}\rangle. 
\end{align}
Here we have adopted the postulate (\ref{mspr1}) that the propagators $G(x, p)$ depend
on the conformal mode $w(x)$.
At short distance when $x\sim y\sim v\sim z$, the vertex correction where the gravitons are exchanged between different legs can be estimated as
\begin{align}
 6 \kappa^2\langle 2w(x) 2w(x)\rangle\frac{1}{P_1^2P_2^2P_3^2P_4^2}. 
\end{align}
We fix the physical momenta $P^2_i$ when we estimate the time evolution of the amplitude. 
There are 6 ways to pair $2w$ out of 4 Green's functions.
After the wave function renormalization we observe this calculation agrees with (\ref{4ECC}).
We can also reproduce (\ref{YECC}) in this way through the conformal mode dependence of the propagators (\ref{mspr1}) and (\ref{mspr2}).
It is our contention that dimensionless couplings acquire time dependence due to soft gravitons. 
We believe that we have clarified its physical mechanism behind it in terms of the effective conformal mode dynamics of the propagators. 
We have observed that these behaviors are consistent with the general covariance if we fix the maximum physical momentum as the UV cut-off.
As these relations follow from the general covariance, we argue that the IR logarithmic corrections are the inevitable consequence of the fundamental symmetry. 

We can also estimate the IR logarithmic corrections to the gravitational couplings at the one-loop level. 
For a conformally flat metric $g_{\mu\nu}=a^2\eta_{\mu\nu}$, the Einstein action is
\begin{align}
\frac{6}{\kappa^2}\int d^4x\ \left( -a\partial_\mu\partial^\mu a
- H^2a^4\right). 
\end{align}
The dS space is the classical solution of this action. Let us assume a more generic 
time dependent background $a$ here. In dS space, we find minimally coupled gravitational modes.
With a generic background $a$, these modes acquire small mass if we assume $a$ is close to
the classical solution.  Such a term gives rise to IR logarithms of the one-loop effective action.
In this way the IR logarithms in the one-loop effective action are estimated as
\begin{align}
\frac{6}{\kappa^2}\int d^4x\ \left( -a\partial_\mu\partial^\mu a
- 2H^2a^4\right)(-\frac{3\kappa^2H^2}{16\pi^2}\log a). 
\end{align}

We find that the inverse of $G$ deceases with time at the one-loop level: 
\begin{align}
\frac{1}{G_\text{eff}} = \frac{1}{G}\big\{1-\frac{3\kappa^2H^2}{16\pi^2}\log a(\tau)\big\}, 
\label{gc1}\end{align}
Furthermore 
we find that $H^2/G$ decreases with time as a soft gravitational effect at the one-loop level:  
\begin{align}
\big(\frac{H^2}{G}\big)_\text{eff} = \big(\frac{H^2}{G}\big) \big\{1-\frac{3\kappa^2H^2}{8\pi^2}\log a(\tau)\big\}. 
\label{gc2}\end{align}
Nevertheless the dimensionless ratio $GH^2$ is the relevant quantity with respect to the cosmological constant problem. 
From (\ref{gc1}) and (\ref{gc2}), we can conclude that soft graviton effects  cancel in this
dimensionless ratio at the one-loop level. 

In fact the above corrections in (\ref{gc1}) and (\ref{gc2}) can be canceled out by shifting the conformal mode of the metric: 
\begin{align}
\Omega  \rightarrow \Omega\{1 +\frac{3\kappa^2H^2}{32\pi^2} \log a(\tau)\}. 
\end{align}
On the other hand, we can show that the mass operators: 
\begin{align}
m^2\Omega^2\tilde{\phi}^2,\ m\Omega\bar{\tilde{\psi}}\tilde{\psi}. 
\end{align}
receive no IR logarithmic corrections at the one-loop level in the Schwinger-Keldysh
formalism.
This fact is consistent with our postulates  (\ref{mspr1}) and (\ref{mspr2}).
Although the above shift removes time dependence in gravitational couplings, it
in turn introduces the IR logarithms into the inertial mass 
\begin{align}
m_\text{eff}=m\big\{1+\frac{3\kappa^2H^2}{32\pi^2} \log a(\tau)\big\}. 
\end{align}
As the physical Newton constant involves $G_\text{eff}m^2=Gm_\text{eff}^2$, we find that it increases with time as
\begin{align}
(Gm^2)_\text{eff} = \big\{1+\frac{3\kappa^2H^2}{16\pi^2} \log a(\tau)\big\}, 
\label{Newtonph}\end{align}
irrespective to our renormalization prescription of the conformal mode.

\section{Gauge dependence}\label{Gauge}
\setcounter{equation}{0}

It is important to investigate the gauge dependence of our obtained results. 
In this section, we adopt the following gauge fixing term with a parameter $\beta$: 
\begin{align}
\mathcal{L}_\text{GF}&=-\frac{1}{2}a^2F_\mu F^\mu, \label{beta}\\
F_\mu&=\beta\partial_\rho h_\mu^{\ \rho}-2\beta\partial_\mu w+\frac{2}{\beta}h_\mu^{\ \rho}\partial_\rho\log a+\frac{4}{\beta}w\partial_\mu\log a. \notag
\end{align}
The gauge condition (\ref{GF}) which is adopted in the preceding sections corresponds with the $\beta=1$ case. 

For an infinitesimal deformation of the gauge parameter: $|\beta^2-1|\ll 1$, the gravitational propagator at the coincident point behaves as 
\begin{align}
\langle (h^{\mu\nu})_+(x)(h^{\rho\sigma})_+(x)\rangle
\to(2-\beta^2)\langle (h^{\mu\nu})_+(x)(h^{\rho\sigma})_+(x)\rangle. 
\label{bgravityp'}\end{align}
Here we have evaluated it perturbatively up to the first order of the gauge deformation: $2-\beta^2=1-(\beta^2-1)$. 

From this fact, we can conclude that the IR logarithmic effects respect the Lorentz invariance for a continuous $\beta$
in scalar and Dirac theory with the wave function renormalization:
\begin{align}
Z_\phi&\simeq 1-(2-\beta^2)\frac{3\kappa^2H^2}{64\pi^2}\log a(\tau), \label{betaZ}\\
Z_\psi&\simeq 1-(2-\beta^2)\frac{3\kappa^2H^2}{256\pi^2}\log a(\tau). 
\end{align} 
It is also the case in gauge theory since there is no wave function renromalization
in a background gauge.

We also find that soft gravitons screen the gauge, quartic and Yukawa couplings in this gauge (\ref{beta}) as 
\begin{align}
e_\text{eff}\simeq e\big\{1-(2-\beta^2)\frac{3\kappa^2H^2}{16\pi^2}\log a(\tau)\big\}, 
\label{betaEGc}\end{align}
\begin{align}
(\lambda_4)_\text{eff}&\simeq \lambda_4\big\{1-(2-\beta^2)\frac{21\kappa^2H^2}{16\pi^2}\log a(\tau)\big\}, 
\label{betaE4c}\end{align}
\begin{align}
(\lambda_Y)_\text{eff}&\simeq \lambda_Y\big\{1-(2-\beta^2)\frac{39\kappa^2H^2}{128\pi^2}\log a(\tau)\big\}. 
\label{betaEYc}\end{align}
As a consequence, we confirm that these effective couplings depend on the gauge parameter. 

Out of these couplings, we can form gauge independent ratios as follows 
\begin{align}
(\lambda_Y)_\text{eff}/\lambda_Y=\big\{(\lambda_4)_\text{eff}/\lambda_4\big\}^\frac{13}{56},\hspace{1em}
e_\text{eff}/e=\big\{(\lambda_4)_\text{eff}/\lambda_4\big\}^\frac{1}{7}. 
\label{rs}\end{align}
We interpret our findings as follows. 
The time dependence of each effective coupling is gauge dependent since there is no unique way to specify the time as it depends on an observer. 
A sensible strategy may be to pick a particular coupling and use its time evolution as a physical time. 
In (\ref{rs}), the coupling of the quartic interaction has been assigned to this role. 
In this setting the relative scaling exponents measure the time evolution of the couplings in terms of a physical time. 
Although the choice of time is not unique, the relative scaling exponents are gauge independent and well defined. 

We can draw an analogy with $2$-dimensional quantum gravity. 
In $2$-dimensional quantum gravity, the couplings acquire nontrivial scaling dimensions due to quantum fluctuations of metric. 
Although the scaling dimension of each coupling is gauge dependent, the ratio of them are gauge independent \cite{KN,KKN}. 
It is because we need to pick a coupling to define a physical scale.
In $4$-dimensional dS space, we can reproduce the IR logarithmic corrections from the conformal mode dependence in the effective propagators: the scalar (\ref{mspr1}) and Dirac field (\ref{mspr2}).
As they follow from the general covariance, we can demonstrate the gauge independence of our such predictions: 
Lorentz invariance and the relative evolution speed of $\lambda_Y$, $\lambda_4$.
It is because the details of the propagator of the conformal mode does not matter. 
In summary we argue that the gauge dependence should be canceled in physical observables such as Lorentz invariance and the ratio of the variation speeds of the couplings.

\section{Self-tuning cosmological constant}\label{Back-reaction}
\setcounter{equation}{0}

In the preceding sections, we have argued that the dimensionless couplings become time dependent and diminish with time in dS space due to soft gravitons.  
The tree level Lagrangian depends on these couplings as 
\begin{align}
\mathcal{L}(\lambda_4, \lambda_Y, e)
=\sqrt{-g} \big[&-\frac{1}{2}g^{\mu\nu}D_{\mu}\phi D_{\nu}\phi+i\bar{\psi}e^\mu_{\ a}\gamma^aD_{\mu}\psi \\
&-\frac{1}{4!}\lambda_4\phi^4 -\lambda_Y\phi\bar{\psi}\psi-\frac{1}{4e^2}g^{\mu\rho}g^{\nu\sigma}F^a_{\mu\nu}F^a_{\rho\sigma}\big]. \notag
\end{align}
The IR effects are local in such a way to make couplings time dependent:
\begin{align}
\mathcal{L}(\lambda_4, \lambda_Y, e)
=\sqrt{-g} \big[&-\frac{1}{2}g^{\mu\nu}D_{\mu}\phi D_{\nu}\phi+i\bar{\psi}e^\mu_{\ a}\gamma^aD_{\mu}\psi \\
&-\frac{1}{4!}\lambda_4(t)\phi^4 -\lambda_Y(t)\phi\bar{\psi}\psi-\frac{1}{4e^2(t)}g^{\mu\rho}g^{\nu\sigma}F^a_{\mu\nu}F^a_{\rho\sigma}\big]. \notag
\end{align}

Such effects may have important physical consequences. 
In an interacting field theory, the cosmological constant is a function of the couplings: $f(\lambda_4, \lambda_Y, e; \Lambda_\text{UV})$ where $\Lambda_\text{UV}$ is the ultra-violet (UV) cut-off. 
As the couplings evolve with time in dS space, the cosmological constant may acquire time dependence: $f(\lambda(t), \lambda_Y(t), e(t); \Lambda_\text{UV})$. 
Here we assume that the UV cut-off $\Lambda_\text{UV}$ is kept fixed. 
Namely the degrees of freedom at the sub-horizon scale are assumed to be constant while the degrees of freedom at the super-horizon scale accumulate with an exponential cosmic expansion. 

Let us consider a time evolution trajectory of the cosmological constant and the couplings. 
Note that the vanishing cosmological constant is the fixed point of the evolution as the couplings stay constant there. 
Even if we start with a theory with a positive cosmological constant, it may be attracted to the fixed point within the domain of the attraction. 
In this way, quantum IR effects may provide a self-tuning mechanism for the cosmological constant. 
We propose it as a simple solution for the cosmological constant problem. 
Its simplicity may underscore the relevance of non-equilibrium physics to this problem. 
In order to provide the existence proof of such a mechanism, we first construct a concrete model with a self-tuning cosmological constant. 

We consider a conformally coupled scalar field with a quartic coupling $\lambda_4$ for simplicity. 
We have argued that gravitons at the super-horizon scale make $\lambda_4$ time dependent: 
\begin{align}
\delta \mathcal{L}_4=-\sqrt{-g}\frac{1}{4!} \lambda_4 (t) \phi^4. 
\label{quartic2}\end{align}
This interaction term contributes to the cosmological constant as a back-reaction. 
It is estimated by taking the vacuum expectation value of this potential term 
\begin{align}
f(\lambda_4 (t);\Lambda_\text{UV})=\frac{1}{4!} \lambda_4 (t) \langle \phi^4 \rangle. 
\end{align}
The leading contribution is quartically UV divergent 
\begin{align}
f(\lambda_4 (t);\Lambda_\text{UV})=\frac{1}{8} \lambda_4 (t)  (\langle \phi^2 \rangle_\text{UV})^2. 
\end{align}
By introducing a physical momentum cut-off $\Lambda_\text{UV}$, we estimate the UV divergence as 
\begin{align}
\langle \phi^2 \rangle_\text{UV}= \frac{\Lambda_\text{UV}^2}{8\pi^2},  
\end{align}
in agreement with (\ref{UVrs}).
We stress here that the time evolution of the couplings is the inevitable consequence of this
postulate.

We can cancel it  by a counter term (bare cosmological constant) at an initial time. 
However we can no longer do so at late times
\begin{align}
 \Delta f(\lambda_4 (t);\Lambda_\text{UV})=\frac{1}{8} \Delta \lambda_4 (t)  (\langle \phi^2\rangle_\text{UV})^2, 
\end{align}
where $\Delta \lambda(t) =\lambda(t)-\lambda$.
In this way the Hubble parameter becomes time dependent:
\begin{align}
\Delta H^2  = \frac{\kappa^2}{48}(\langle \phi^2 \rangle_\text{UV})^2\Delta \lambda_4 (t) = DM_P^2\Delta \lambda_4 (t), 
\end{align}
where the coefficient $D$ is
\begin{align}
D=\frac{\pi}{3}\big(\frac{1}{8\pi^2}\frac{\Lambda_\text{UV}^2}{M_P^2}\big)^2. 
\end{align}
The essential point here is that the bare action is assumed to be generally covariant 
and hence dS invariant. 
The symmetry breaking  effect comes from IR effects. 
With this assumption, we can estimate the time evolution of the cosmological constant even if we cannot predict its initial value. 

The evolution trajectory of the cosmological constant is a straight line in $(\lambda_4, H^2)$ plane.
This trajectory in $( \lambda_4, H^2 )$ plane is attracted to a fixed point on the horizontal line with $H^2=0$ 
as $\lambda_4$ is decreased by $\Delta \lambda_4 = H^2/(DM_P^2)$. 
We should emphasize that this statement is gauge independent. 
\begin{align}
\parbox{\lambdaHl}{\usebox{\lambdaH}}
\end{align}

Furthermore let us see the implication of this model for Dark Energy. 
Since $H^2/M_P^2 \sim 10^{-120}$ at present,
$\lambda_4$ changes little in this process as long as $D\sim (\Lambda_\text{UV}/M_P)^4$ is not so small. 
This condition is easily satisfied with $\Lambda_\text{UV} > m_{\nu}$. 

As seen in (\ref{4ECC}), the effective coupling of the quartic interaction decreases with time as 
\begin{align}
\Delta \lambda_4 = - \lambda_4 C\frac{H^2}{M_P^2}\log a(t),\hspace{1em}C=\frac{21}{\pi}. 
\end{align}
Through the effective coupling, the Hubble parameter acquires the following time dependence 
\begin{align}
\Delta H^2 = -\lambda_4 CD H^2 \log a(t).  
\end{align}
As $\lambda_4$ changes little, we can ignore the time dependence of $\lambda_4$ in this equation. 
In this way, we find 
\begin{align}
H^2 \sim a(t)^{-n},\hspace{1em}n=\lambda_4 CD. 
\label{index}
\end{align}
As is well known, $n=3$ for dark matter and $n=0$ for cosmological constant.
The scaling index $n$ is small $n < 10^{-3}$ as long as we keep the UV cut-off of this model $\Lambda_\text{UV} < M_P$. 
The important point is that $n$ is non-vanishing unlike a true cosmological constant. 
The Hubble parameter $H^2$ is self-tuned to zero at late times in this model. 

The theoretical issue we need to clarify here is that the coefficient $C$ depends on a gauge parameter $\beta$ as
\begin{align}
C_\beta=\frac{21}{\pi}(2-\beta^2), 
\end{align}
where we have evaluated it up to the first order of the gauge deformation in a similar way to (\ref{bgravityp'}). 
As seen in (\ref{betaE4c}), it originates in the fact that the evolution speed of the coupling is gauge dependent:
\begin{align}
\Delta \lambda_4 = - \lambda_4C_\beta\frac{H^2}{M_P^2}\log a(t).  
\end{align}
We observe that  this factor can be absorbed by the reparametrization of the scale factor up to $\mathcal{O}(\beta^2-1)$ 
\begin{align}
(2-\beta^2)\log a(t) \sim \frac{1}{\beta^2}\log a(t)= \log a'(t'). 
\end{align}
Of course it is due to the fact that time is an observer dependent quantity.
We hence argue that such an ambiguity must be resolved by specifying the coordinates of the observer at the quantum level. 
In this  procedure the symmetry may play an important role. 
For example the $\beta =1$ case corresponds to the most symmetric space-time since the gravitational propagators possess $SO(3)$ symmetry as seen in (\ref{minimally}).
Namely  the scalar, vector and tensor modes of $\tilde{h}^i_{\ j}$ are degenerated in this gauge. 

We remark that use of the dimensional regularization to remove power divergences are not allowed in our rule.
It is because such a subtraction does not correspond to dS invariant counter terms.

We also mention that supersymmetry (SUSY) does not remove this effect either. 
SUSY removes quartic divergences of the vacuum energy as long as quartic, Yukawa and gauge coupling are related in a specific way. 
However as seen in (\ref{EGc}), (\ref{4ECC}) and  (\ref{YECC}), they evolve differently with time in dS space. 
So SUSY relations of the couplings are split with time and quartic divergences of the vacuum energy no longer cancel. 
It is also clear that quadratic divergences also remain and they split mass degeneracy among SUSY multiplets. 

\section{Splitting SUSY}\label{SUSY}
\setcounter{equation}{0}

In this section, we investigate SUSY splitting effect in dS space in more detail.
Let us consider a Wess-Zumino model with a superpotential
\begin{align}
W=\frac{1}{3}g \Phi ^3, 
\end{align}
where $\Phi$ denotes a chiral super field. 
In terms of component fields: complex scalar and Weyl spinor\footnote{Our convention is $\psi\psi=\epsilon^{\alpha\beta}\psi_\alpha\psi_\beta$, $\epsilon^{12}=-\epsilon^{21}=1$, $\epsilon^{11}=\epsilon^{22}=0$.}, 
the interaction term is expressed as 
\begin{align}
-g_4^2|\phi^2|^2-g_3(\phi\psi\psi+\text{h.c.} ), 
\end{align}
where $g_4=g_3=g$.

As seen in Section \ref{Interaction}, each coupling is screened by soft gravitons as 
\begin{align}
\Delta g_4^2 = - \frac{21}{16\pi^2} {\kappa^2 H^2} \log a(t) g^2, 
\end{align}
\begin{align}
\Delta g_3^2 = - \frac{39}{64 \pi^2} {\kappa^2H^2} \log a(t) g^2. 
\end{align}
As a consequence, the UV-IR mixing effect on the Hubble parameter is evaluated as
\begin{align}
\Delta H^2 &= 16DM_P^2( \Delta g_4^2 - \Delta g_3^2 ) \\
 &=-\frac{180}{\pi}DH^2\log a(t) g^2. \notag
\end{align}

The scalar mass term  is generated as
\begin{align}
\Delta m_\phi^2 &= 4 \langle \phi^2 \rangle_\text{UV} ( \Delta g_4^2 - \Delta g_3^2 ) \\
&= -\frac{45}{\pi}E H^2 \log a(t) g^2, \notag
\end{align}
where $E=\Lambda_\text{UV}^2/ 8\pi^2 M_P^2$.
It is of order Hubble scale $H^2$.
The fermion mass term is not perturbatively generated due to the chiral symmetry.

Note that the scalar field becomes tachyonic $m_\phi^2 < 0$ if we start with a massless super multiplet.
Therefore the scalar field acquires the vacuum expectation value $|\phi|=\sqrt{2}|m_\phi|/g$.
Since we have a Mexican hat type potential, we obtain a Nambu-Goldstone boson.
The mass of a Higgs type boson is $\sqrt{2}|m_\phi|$. The fermions acquire the identical mass.

We may draw a lesson from this analysis of a simple SUSY model.
In dS space, SUSY degeneracy in the mass spectrum is lifted. The massless particles
acquire the mass of order Hubble scale unless they are protected by symmetries.
In other words, the gauge bosons with gauge symmetry and fermions with chiral
symmetry remain massless unless the relevant symmetry is spontaneously broken.

The origin of large structure of the Universe is argued to be generated during cosmic inflation with the potential $V \sim (10^{16} \text{GeV})^4 (r/0.1)$. 
In the inflation era, the Hubble scale is $H/M_P \sim 10^{-5} $ unless the tensor to scalar ratio $r$ is much smaller than the current upper bound of $0.1$. 

It is likely that SUSY is required to make a consistent theory of quantum gravity such as string theory.
However our analysis indicates that inflation splits SUSY with SUSY breaking scale of $H \sim 10^{-5} M_P$.
A possible way out to have TeV scale SUSY is to make $r$ ratio very small as $10^{-20}$.  
We may thus conclude:
If Planck observes primordial tensor modes, LHC will find no SUSY and vise versa.

It is certainly desirable to investigate this issue in a UV finite quantum gravity such as string theory.
This mechanism could be investigated nonperturbatively by Minkowski version of IIB matrix model
\cite{KNT}.
It is interesting to note that an exploratory investigation indicates the beginning of the
universe with Inflation.
We could also list possibly relevant questions:

\begin{itemize}
\item
Can we understand Higgs mass $m_H$ from this perspective?
\item
What are the implications to inflation theory?
\end{itemize}

\section{Diminishing cosmological constant in 2d gravity}\label{Liouville}
\setcounter{equation}{0}

In this section we investigate the quantum IR effects in a solvable model: 2-dimensional (2d) quantum gravity. 
We choose a conformal gauge 
\begin{align}
g_{\mu\nu}=e^{\phi}\hat{g}_{\mu\nu}, 
\end{align}
where $\hat{g}_{\mu\nu}$ is a background metric.
The effective action for the conformal mode $\phi$ is the Liouville action:
\begin{align}
 \int \sqrt{-\hat{g}}d^2x \big[-\frac{25-c}{96\pi}(\hat{g}^{\mu\nu}\partial_{\mu}\phi\partial_{\nu}\phi
+2\phi\tilde{R})-\Lambda e^{(1+\frac{\gamma}{2})\phi}\big]. 
\end{align}
Here $c$ denotes the central charge of the matter coupled to 2d quantum gravity.
In the free field case, $c$ counts massless scalars and fermionic fields: $c=N_s+N_f/2$.
We consider the semiclassical regime: $c > 25$ where the metric for the conformal mode is negative and hence time-like. 
It is the identical feature with 4d Einstein gravity. 
In the above expression $\Lambda$ is the cosmological constant, $e^{(1+\frac{\gamma}{2})\phi}$ is a renormalized cosmological constant operator and $\gamma$ denotes the anomalous dimension. 

The equation of motion with respect to $\phi$ is given by 
\begin{align}
\frac{25-c}{48\pi}\nabla^2\phi-\Lambda e^{\phi}=0, 
\end{align}
where we put $\gamma =0$.
Furthermore the equation of motion with respect to $h^{\mu\nu}$ is 
\begin{align}
\frac{25-c}{48\pi}\left(\nabla_{\mu}\phi\nabla_{\nu}\phi-2\nabla_{\mu}\nabla_{\nu}\phi\right)=\nabla_{\mu}f\nabla_{\nu}f. 
\end{align}
where $f$ denotes a free scalar field.
The classical solution of the Liouville theory is 2d dS space: 
\begin{align}
e^{\phi_c}=\big(\frac{1}{-H\tau}\big)^2,\hspace{1em}H^2=\frac{24\pi}{c-25}\Lambda. 
\end{align}
We identify cosmic time with the  classical solution for the conformal mode $\phi_c (t)=2Ht$.
On the other hand if the cosmological constant can be neglected, we also have a solution with a non-trivial free matter field $f$: 
\begin{align}
\phi_c = A\tau,\hspace{1em}f_c=A\sqrt{\frac{25-c}{48\pi}}\tau. 
\end{align}
where $A$ is an arbitrary constant. This is a 2d Friedmann spacetime. 
This solution should go over to the 2d dS space solution when the cosmological constant becomes dominant. 

To renormalize the cosmological constant operator to the leading order in $1/( c-25)$, 
we need to consider the quantum fluctuation of the bare cosmological constant operator: 
\begin{align}
\langle e^{\phi} \rangle \sim e^{\phi_c(t)+\frac{1}{2}\langle \phi^2 \rangle}. 
\end{align}
Here $\phi_c(t)$ denote a classical solution while
\begin{align}
\langle \phi^2 \rangle=-\frac{24}{c-25}\int_{P_\text{min}}^{P_\text{max}} \frac{dP}{P}. 
\end{align}
The scalar propagator is both UV and IR divergent in 2-dimension. 
In this integral with respect to physical momentum, we fix the UV cut-off $P_\text{max}$ while we identify the IR cut-off as $P_\text{min}=L/a(t)$. 
Here $a(t)=e^{\phi_c(t)/2}$ is the scale factor of the Universe and $L$ is the initial size of the universe. 
In this way the quantum IR fluctuation grows as the Universe expands:
\begin{align}
\langle \phi^2 \rangle\sim-\frac{12}{c-25}\phi_c (t)\ \Rightarrow\ \langle e^\phi\rangle \sim e^{(1-\frac{6}{c-25})\phi_c (t)}. 
\end{align}
We have thus found that the effective cosmological constant diminishes as the Universe expands: 
\begin{align}
H^2_\text{eff} \sim H^2e^{-\frac{6}{c-25}\phi_c (t)}\sim H^2a(t)^{-\frac{12}{c-25}}. 
\end{align}
The important point here is that the quantum IR effect is time dependent and hence cannot be subtracted by a dS invariant counter term.

The scaling dimension of the cosmological constant operator is
\begin{align}
1+\frac{\gamma}{2}=\frac{2}{1+\sqrt{1+\frac{24}{c-25}}}\sim 1-\frac{6}{c-25} + \cdots. 
\label{anmd}\end{align}
We have reproduced the exact result to the leading order in a simple argument.
The exact expression shows that the anomalous dimension $\gamma$ is negative in the semiclassical regime $c > 25$. 
Therefore we can conclude that the quantum IR effects make  cosmological constant diminish with time beyond perturbation theory. 
As the quantum IR effects in 4d dS space can be understood in terms of
the conformal mode as well, this model illustrates how the couplings acquire the time  dependence.

The effective Newton's constant also decreases with time
\begin{align}
\frac{1}{G_\text{eff}}=\frac{1}{G} +\frac{c-25}{48\pi}\phi_c (t). 
\label{topo}
\end{align}
In 2-dimension this coupling is topological.
For $c<1$, (\ref{anmd}) and (\ref{topo}) represent the double scaling relation between these couplings.
In such a situation the both couplings grow in the IR limit and we need to fine tune them.

So far we have assumed that the matter system is at the critical point, namely conformally invariant.
In a more generic situation, the central charge $c$ is known to be a decreasing function with respect to the
IR cut-off and hence time $\phi_c (t)$. 
For example a nonlinear sigma model may develop a mass gap. 
In such a situation, the number of massless scalar fields decreases.
This effect may enhance the magnitude of the anomalous dimension and 
the screening effect of the cosmological constant. 

\section{Conclusion}\label{Conclusion}
\setcounter{equation}{0}

The dS symmetry may be broken due to growing IR effects.
Since there exist massless minimally coupled modes in the metric fluctuations,
such an effect may be a ubiquitous feature of quantum gravity in dS space.
In particular we have found that the couplings acquire time dependence
and evolve logarithmically with the scale factor of the Universe. 
We thus propose that the relative evolution speeds of the couplings are the physical observables in dS space. 

We observe these effects can be explained by the nontrivial conformal mode dependence of the operators.
We recall here the analogous renormalization takes place in 2d quantum gravity.
There the scaling dimensions of the local operators and equivalently couplings to them are renormalized by the scale invariant quantum fluctuation of the metric.
We have recalled that the cosmological constant is screened by IR quantum fluctuations in the semiclassical regime.
We believe therefore analogous renormalization takes place in 4d dS space due to the scale invariant quantum IR fluctuations of the metric. 
We argue that the IR logarithmic effects originate from the conformal mode dependence in the effective propagators of the scalar and Dirac fields.
In fact we have shown such a quantum effect is the inevitable consequence of the general covariance. 
Thus we have made a strong case for the validity of our claim when these fields are involved.

The magnitude of this effect is of $O(H^2/M_P^2)$.
Since this factor is very small $(10^{-120})$ now, it seems impossible to detect such an effect at first sight.
However this factor may be compensated by UV divergences in the case of the cosmological constant which is quartically divergent with respect to the UV cut-off.
If so, such an effect could lead to very significant consequences:
\begin{itemize}
\item
{Dark energy may decay with a small but finite $n < 10^{-3}$.}
\item
Inflation may split SUSY in the early universe.
\end{itemize}

Let us recall BPHZ renormalization procedure.
We assume that the theory is renormalized at the $n$-th order with appropriate counter terms.
In the next order, the new divergences occur as over-all divergences since sub-divergences are assumed to be subtracted.
If the couplings are time independent, we can renormalize these new local divergences by covariant counter terms. 
In this way renormalizability implies the absence of UV-IR mixing effect and vise versa. 
On the other hand, we cannot do so if the couplings become time dependent.
The UV-IR mixing effect could arise as the incomplete cancellation of the divergences due to time evolution of the couplings.


So UV-IR mixing effects could occur in non-equilibrium and non-renormalizable field theory.
Although our reasoning is different, we agree with Polyakov that the cosmological constant problem may very well be of this nature \cite{Polyakov}.

Let us consider the density fluctuation in the cosmic microwave background (CMB) due to scalar modes.
The standard argument relates it to the two point function of the minimally coupled scalar field (inflaton) at the horizon crossing as it is identified with the curvature perturbation which remains constant
after the horizon crossing in the comoving gauge \cite{Mukhanov1990,Maldacena2002}.
We may be able to estimate the scalar and tensor perturbations from the sub-horizon theory side by following it up to the horizon crossing.
We have shown that the leading IR effects can be absorbed into the renormalization of the fields and couplings inside the cosmological horizon. 
We expect that this picture still holds when the space time is approximately dS such as in the inflation era.
We hence believe that our results have an implication for the super-horizon soft graviton loop effect on CMB. 
If so, it shows that the leading IR effects on the scalar two point functions due to soft gravitons are absent as they can be renormalized away. 
The time evolution of the couplings modifies slow roll parameters $\epsilon$ and $\eta$ in general. 

The curvature and tensor perturbations are shown to stay constant after the horizon crossing in the comoving gauge at the tree level in a single field inflation.
It is because they approach pure gauge.
The nontrivial potential for them are forbidden due to general covariance.
The issues on the higher loop corrections are reviewed in \cite{Tanaka}.
The IR logarithmic corrections influence the CMB spectrum by making the microscopic parameters of the inflation theory time dependent.
We believe that it is important to understand them when we investigate microscopic physics behind inflation. 

We illustrate one of such a possibility.
The self-tuning effect of the cosmological constant due to UV-IR mixing could alter the classical slow roll picture. 
The slow roll parameters are
\begin{align}
\epsilon =\frac{1}{16\pi}(\frac{M_PV'}{V})^2\sim 4\pi\frac{\dot{\phi}^2}{H^2}\frac{1}{M_P^2},\hspace{1em}\eta=\frac{1}{8\pi}\frac{M_p^2V''}{V}. 
\end{align}
We note that the slow roll parameter $\epsilon$ acts like $n/2$ in (\ref{index})
\begin{align}
H^2 \sim a(t)^{-2\epsilon}.
\end{align}
So the quantum effects could significantly alter the classical slow roll picture when $\epsilon$ becomes as small as $n$. 
In the case of the single field inflation, $n$ may be small as $\lambda_4$ is strongly constrained by the observation. 
In a generic case, $n$ could be as large as $10^{-3}$ since it could arise from different fields rather than the inflaton.
The quantum effect alters the classical slow roll picture in general if the tensor-to-scalar ratio $r$ is as small as $0.01$. 
Such a parameter region will be explored soon. 

\section*{Acknowledgment}
This work is supported in part by the Grant-in-Aid for Scientific Research 
from the Ministry of Education, Science and Culture of Japan, 
and the National Research Foundation of Korea (NRF) grant funded by the Korea government (MSIP) 
through Seoul National University with grant numbers 2005-0093843, 2010-220-C00003 and 2012K2A1A9055280. 
We thank S. Iso, J. Nishimura, H. Kodama and Soo-Jong Rey for discussions. 


\end{document}